\documentclass[final,5p,times,twocolumn]{elsarticle} 

\usepackage{amssymb}
\usepackage{amsmath}

\usepackage[caption=false,font=footnotesize]{subfig}
\usepackage{booktabs}				
\usepackage{array}					
\usepackage{multirow}				
\usepackage{siunitx}				
\usepackage{flafter}				
\usepackage{comment}				
\usepackage[usenames,dvipsnames]{xcolor}	
\usepackage{flafter}				
\usepackage{lipsum} 				
\usepackage{setspace}               
\usepackage{marginnote}             
\usepackage{multicol}

\usepackage{amssymb}
\usepackage{soul}
\usepackage{url}
\usepackage{epstopdf}
\usepackage{rotating}
\usepackage{placeins}
\usepackage{pifont}
\usepackage{threeparttable}
\usepackage{makecell}
\biboptions{numbers,sort&compress}
\usepackage{hyperref}	
\usepackage{breakurl}
\usepackage{cleveref}	
\usepackage{amsmath, colortbl}
\usepackage{siunitx}
\usepackage{xr}
\usepackage{nomencl}

\makenomenclature
\usepackage{etoolbox}
\usepackage[most]{tcolorbox}
\makeatletter
\newcommand{\inputgeneratednomenclature}[2][\nomlabelwidth]{%
  \begingroup
  \nom@tempdim=#1\relax
  \input{#2}%
  \endgroup
}
\makeatother
\newtcolorbox{NomBox}{
  enhanced,
  breakable,
  colback=white,
  colframe=black,
  boxrule=0.6pt,
  arc=0pt,
  left=4pt,right=4pt,top=4pt,bottom=4pt,
  width=\columnwidth
}

\renewcommand{\nomgroup}[1]{%
  \item[\bfseries
    \ifstrequal{#1}{A}{Sets and indices}{%
    \ifstrequal{#1}{B}{Parameters}{%
    \ifstrequal{#1}{C}{Decision variables}{}}}]%
}

\nomenclature[A]{$t\ \epsilon\ \mathcal{T}$}{Time steps within the day.}
\nomenclature[A]{$v\ \epsilon\ \mathcal{V}$}{Set of plugged-in electric vehicles.}
\nomenclature[A]{$e\ \epsilon\ \mathcal{S}_v$}{Session-end time indices for vehicle $v$.}
\nomenclature[A]{$\mathcal{F}_{\mathrm{LinDistFlow}}$}{Feasible set defined by the linearized DistFlow power-flow equations.}
\nomenclature[A]{$g(\cdot)$}{Vector-valued function representing grid operational limit constraints (e.g., voltage and branch loading).}
\nomenclature[B]{$\Delta t$}{Time-step length (h).}
\nomenclature[B]{$\bar P_t$}{CP profile (kW).}
\nomenclature[B]{$U_t,\,L_t$}{Upper/lower bounds for total EV charging (kW) at time step $t$.}
\nomenclature[B]{$\eta_v$}{Charging efficiency of EV $v$.}
\nomenclature[B]{$P^{\max}_v$}{Nameplate power of EV $v$’s charger (kW).}
\nomenclature[B]{$SOE^{\min}_v$}{SoE min. bound of EV $v$ (kWh).}
\nomenclature[B]{$SOE^{\max}_v$}{SoE max. bound of EV $v$ (kWh).}
\nomenclature[B]{$E^{\mathrm{trip}}_{v,e}$}{Energy required for the next trip of EV $v$ at session end $e$ (kWh).}
\nomenclature[B]{$SOE^{\mathrm{unctrl,end}}$}{Aggregate end-of-day SoE under uncontrolled charging (kWh).}
\nomenclature[B]{$\beta$}{Weighting parameter to prioritize grid constrains over cost minimization.}
\nomenclature[B]{$c_{el,t}$}{Cost of electricity at time $t$ (CHF/kWh).}
\nomenclature[B]{$\bar C^{\mathrm{PV}}$}{Maximum total PV curtailment over the analyzed period, obtained from the CP simulation (kWh).}
\nomenclature[B]{$\rho$}{Weighting parameter to prioritize profile adherence over grid constraints.}
\nomenclature[B]{$\bar G^{\mathrm{PV}}_t$}{Available PV generation at time $t$ (kW).}

\nomenclature[C]{$p_{v,t}$}{Charging power of EV $v$ at time $t$ (kW).}
\nomenclature[C]{$SOE_{v,t}$}{State of energy of EV $v$ at time $t$ (kWh).}
\nomenclature[C]{$z_t$}{Absolute deviation from CP profile at time $t$ (kW).}
\nomenclature[C]{$D_{\mathrm{res},t}$}{Total residual demand, after local PV generation, at time $t$ (kWh).}
\nomenclature[C]{$G_{\mathrm{res},t}$}{Total residual generation, after local consumption, at time $t$ (kWh).}
\nomenclature[C]{$c^{\mathrm{PV}}_{t}$}{PV power curtailment at time $t$ (kW).}
\nomenclature[C]{$P_t$}{Vector of active power flows on distribution lines at time $t$ (kW).}
\nomenclature[C]{$Q_t$}{Vector of reactive power flows on distribution lines at time $t$ (kVAr).}
\nomenclature[C]{$V_t$}{Vector of squared nodal voltage magnitudes at time $t$ (V $^2$).}
\nomenclature[C]{$s_t$}{Nonnegative slack variable aggregating violations of grid operational constraints across nodes and branches at time $t$.}
\nomenclature[C]{$G^{\mathrm{PV}}_t$}{PV generation injected into the grid at time $t$ (kW).}
\journal{arXiv}
\begin{document}

\begin{frontmatter}

\title{Does Central Planning Fail Locally? Evaluating EV Charging Flexibility Optimization Across Grid Levels}

\author[RRE]{Ambra Van Liedekerke\corref{contrib}}
\author[RRE]{Lorenzo Zapparoli\corref{contrib}}
\author[PSL]{María Parajeles Herrera\corref{contrib}}
\author[RRE]{Blazhe Gjorgiev}
\author[PSL]{Gabriela Hug}
\author[RRE]{Giovanni Sansavini\corref{cor1}} \ead{sansavig@ethz.ch}

\affiliation[RRE]{organization={Reliability and Risk Engineering Laboratory, Institute of Energy and Process Engineering, Department of Mechanical and Process Engineering, ETH Zurich},
            city={Zurich},
            country={Switzerland}}
\affiliation[PSL]{organization={Power System Laboratory, Institute of Information Technology and Electrical Engineering, ETH Zurich},
            city={Zurich},
            country={Switzerland}}
\cortext[cor1]{Corresponding author.}
\cortext[contrib]{Authors contributed equally.}

\begin{abstract} 
Electric vehicles (EVs) are a key enabler of global decarbonization, and their charging flexibility is crucial for integrating high shares of renewable energy into future power systems.
Research on EV charging flexibility in future power systems has largely focused on either the transmission or the distribution level, with limited integration between the two. 
However, these two scales cannot be separated: EV charging optimized at the national transmission level for system-wide planning is ultimately activated in local distribution grids, where it may cause voltage and loading issues. 
This study bridges that gap by evaluating EV charging flexibility across both grid levels, using two different control approaches.
First, a top-down approach disaggregates the optimal charging demand obtained from a centralized planner at the transmission grid down to the distribution grid, explicitly accounting for grid constraints and individual driving requirements.
Second, a bottom-up approach optimizes EV charging locally under distribution-grid constraints using electricity price signals from the centralized planning model.
Controlled charging from the top-down and bottom-up approaches is then compared with uncontrolled charging.
Results show that the top-down approach produces EV charging schedules that are compatible with individual driving behavior and reduce grid violations compared to uncontrolled charging.
The bottom-up approach identifies an optimal profile similar to the centrally optimized profile, showing that electricity price signals are a suitable coordination mechanism.
These findings suggest that centrally optimized EV charging remains effective when implemented at the distribution level, in systems with high DER penetration.

\end{abstract}


\end{frontmatter}

\section{Introduction} 
\label{intro}

Electric vehicles are a key enabler of global decarbonization~\cite{IEA_outlook2024}.
Transport electrification significantly reduces total energy consumption and CO$_2$ emissions, provided it is supported by the expansion of renewable electricity generation~\cite{Athanasopoulou2018, Zhao2023}.
    At the same time, widespread EV adoption presents new challenges, increasing overall electricity demand and system peak load~\cite{IEA_outlook2024}, and impacting voltage profiles in distribution grids~\cite{Soofi2022}.
    Uncontrolled EV deployment could raise peak demand by 35–51\% across European countries~\cite{Mangipinto2022},
    necessitating substantial investment in generation capacity and distribution grids ~\cite{Crozier2020}.
    However, since EVs remain parked approximately 96\% of the time~\cite{Razipour2019}, they can provide significant demand-side flexibility through controlled charging strategies and bidirectional charging, commonly referred to as vehicle-to-grid (V2G).
    Notably, controlled EV charging alone accounts for approximately half of the household flexibility potential~\cite{Harder2020}.
    Beyond controlled charging, V2G enables power injection into the grid, offering even greater operational flexibility, though its benefits compared to unidirectional controlled charging remain highly context-dependent~\cite{Brodnicke2025}.
    Overall, EV charging flexibility is considered essential for the reliable integration of high shares of renewable energy into a fossil-free power system~\cite{Kempton2005}.

At the scale of centralized operation and generation capacity planning, EV charging flexibility is recognized as a key enabler of the low-carbon transition, as it reduces reliance on conventional generation and storage~\cite{Owens2022, Crozier2020} and supports renewable integration~\cite{Taljegard2019}.
    In France alone, EV charging flexibility could reduce annual operational costs by up to €1.1 billion and CO$_2$ emissions by 3.2 Mt within the 2040 European interconnected grid~\cite{Lauvergne2022}.
    Similar findings emerge for a UK case study, where the value of EV charging flexibility increases with higher levels of transport electrification~\cite{Ramirez2016}.
    On a European scale, EV charging flexibility leads to annual system cost savings between €19 and 42 billion and substantially reduces infrastructure requirements~\cite{Sanvito2026}.
    Further evidence highlights additional benefits, such as lower electricity prices~\cite{Wolinetz2018}, and significant reductions in grid congestion in Germany’s 2030 transmission grid~\cite{Staudt2018, Ruppert2023}.
    High levels of EV charging flexibility in the 2020–2050 EU pathway can decrease the need for transmission expansion, reduce system costs, and cut CO$_2$ emissions~\cite{Gunkel2020, Sanvito2026}.

\begin{NomBox}
  \small
  \setlength{\parindent}{0pt}
  \setlength{\parskip}{0pt}
  \setlength{\nomitemsep}{0.1\baselineskip}

  \makeatletter
  \nom@tempdim=0.35\columnwidth
  \makeatother

  \begin{thenomenclature}
    \nomgroup{A}
      \item [{$\mathcal{F}_{\mathrm{LinDistFlow}}$}]\begingroup Feasible set defined by the linearized DistFlow power-flow equations.\nomeqref {0}\nompageref{1}
      \item [{$e\ \epsilon\ \mathcal{S}_v$}]\begingroup Session-end time indices for vehicle $v$.\nomeqref {0}\nompageref{1}
      \item [{$g(\cdot)$}]\begingroup Vector-valued function representing grid operational limit constraints (e.g., voltage and branch loading).\nomeqref {0}\nompageref{1}
      \item [{$t\ \epsilon\ \mathcal{T}$}]\begingroup Time steps within the day.\nomeqref {0}\nompageref{1}
      \item [{$v\ \epsilon\ \mathcal{V}$}]\begingroup Set of plugged-in electric vehicles.\nomeqref {0}\nompageref{1}
    \nomgroup{B}
      \item [{$\bar C^{\mathrm{PV}}$}]\begingroup Maximum total PV curtailment over the analyzed period, obtained from the CP simulation (kWh).\nomeqref {0}\nompageref{1}
      \item [{$\bar G^{\mathrm{PV}}_t$}]\begingroup Available PV generation at time $t$ (kW).\nomeqref {0}\nompageref{1}
      \item [{$\bar P_t$}]\begingroup CP profile (kW).\nomeqref {0}\nompageref{1}
      \item [{$\beta$}]\begingroup Weighting parameter to prioritize grid constrains over cost minimization.\nomeqref {0}\nompageref{1}
      \item [{$\Delta t$}]\begingroup Time-step length (h).\nomeqref {0}\nompageref{1}
      \item [{$\eta_v$}]\begingroup Charging efficiency of EV $v$.\nomeqref {0}\nompageref{1}
      \item [{$\rho$}]\begingroup Weighting parameter to prioritize profile adherence over grid constraints.\nomeqref {0}\nompageref{1}
      \item [{$c_{el,t}$}]\begingroup Cost of electricity at time $t$ (CHF/kWh).\nomeqref {0}\nompageref{1}
      \item [{$E^{\mathrm{trip}}_{v,e}$}]\begingroup Energy required for the next trip of EV $v$ at session end $e$ (kWh).\nomeqref {0}\nompageref{1}
      \item [{$P^{\max}_v$}]\begingroup Nameplate power of EV $v$’s charger (kW).\nomeqref {0}\nompageref{1}
      \item [{$SOE^{\mathrm{unctrl,end}}$}]\begingroup Aggregate end-of-day SoE under uncontrolled charging (kWh).\nomeqref {0}\nompageref{1}
      \item [{$SOE^{\max}_v$}]\begingroup SoE max. bound of EV $v$ (kWh).\nomeqref {0}\nompageref{1}
      \item [{$SOE^{\min}_v$}]\begingroup SoE min. bound of EV $v$ (kWh).\nomeqref {0}\nompageref{1}
      \item [{$U_t,\,L_t$}]\begingroup Upper/lower bounds for total EV charging (kW) at time step $t$.\nomeqref {0}\nompageref{1}
    \nomgroup{C}
      \item [{$c^{\mathrm{PV}}_{t}$}]\begingroup PV power curtailment at time $t$ (kW).\nomeqref {0}\nompageref{1}
      \item [{$D_{\mathrm{res},t}$}]\begingroup Total residual demand, after local PV generation, at time $t$ (kWh).\nomeqref {0}\nompageref{1}
      \item [{$G^{\mathrm{PV}}_t$}]\begingroup PV generation injected into the grid at time $t$ (kW).\nomeqref {0}\nompageref{1}
      \item [{$G_{\mathrm{res},t}$}]\begingroup Total residual generation, after local consumption, at time $t$ (kWh).\nomeqref {0}\nompageref{1}
      \item [{$P_t$}]\begingroup Vector of active power flows on distribution lines at time $t$ (kW).\nomeqref {0}\nompageref{1}
      \item [{$p_{v,t}$}]\begingroup Charging power of EV $v$ at time $t$ (kW).\nomeqref {0}\nompageref{1}
      \item [{$Q_t$}]\begingroup Vector of reactive power flows on distribution lines at time $t$ (kVAr).\nomeqref {0}\nompageref{1}
      \item [{$s_t$}]\begingroup Nonnegative slack variable aggregating violations of grid operational constraints across nodes and branches at time $t$.\nomeqref {0}\nompageref{1}
      \item [{$SOE_{v,t}$}]\begingroup State of energy of EV $v$ at time $t$ (kWh).\nomeqref {0}\nompageref{1}
      \item [{$V_t$}]\begingroup Vector of squared nodal voltage magnitudes at time $t$ (V $^2$).\nomeqref {0}\nompageref{1}
      \item [{$z_t$}]\begingroup Absolute deviation from CP profile at time $t$ (kW).\nomeqref {0}\nompageref{1}
    
    \end{thenomenclature}
\end{NomBox}

EV charging flexibility is also recognized as a key enabler of the energy transition at the distribution grid level,
    where it can increase residential photovoltaic (PV) integration~\cite{Schwarz2020} and enhance the hosting capacity of EVs themselves~\cite{Kamruzzaman2019}.
    Moreover, EV charging flexibility can alleviate congestion in distribution grids~\cite{Sundstrom2012} and reduce the need for costly grid expansion~\cite{Hemmati2020, spiliotis2016}.
    Beyond these impacts, peak shaving~\cite{Nespoli2023}, provision of ancillary services~\cite {Sortomme2012}, and voltage control~\cite{DeSantis2023} are additional benefits.

Fully exploiting EV flexibility for operational purposes at both scales, however, requires effective coordination between transmission system operators (TSOs) and distribution system operators (DSOs)~\cite{Grottum2019}. 
This complex challenge remains an active area of research.
A primary focus in this area is the design of TSO-DSO coordination frameworks capable of managing large numbers of flexible units distributed across grids~\cite{Venegas2025, Shao2016}, 
while safeguarding the mobility needs of EV owners~\cite{Wang2022}.
The research spans both advanced optimization methods and the design of market mechanisms to facilitate coordinated operational planning and the provision of key services, including voltage control, congestion management, frequency regulation, system restoration, and ancillary services~\cite{Zahraoui2025}.

Despite extensive research, most studies address EV charging flexibility from either the transmission or the distribution perspective, with limited integration of both viewpoints.
In particular, it remains unclear if EV charging flexibility optimized for centralized system expansion planning is compatible with distribution grid operations and individual driving requirements.
Conversely, it remains unclear how centrally optimized EV charging compares with locally optimized charging in response to electricity price signals.
Research on TSO–DSO coordination partially bridges these perspectives; however, it primarily focuses on operational and market design aspects and is usually limited to medium-scale distributed energy resources (DERs), hence, only considering medium-voltage grids~\cite{Givisiez2020}.
Moreover, spatio-temporal fidelity and end-user behavior, such as the individual EV driving patterns, remain overlooked~\cite{Muratori2020, Mohanty2022}, despite their strong influence on charging flexibility availability~\cite{Panda2022}.
It remains thus unclear whether the transmission and the distribution perspectives converge toward similar charging patterns or produce conflicting outcomes at the low-voltage level.

To address these gaps, this paper investigates controlled EV charging taking into account both the transmission and distribution system levels. 
Specifically, we analyze how centralized controlled EV charging optimization embedded in centralized system operation and generation capacity planning translates into impacts on distribution grids. 
We further compare the outcomes of decentralized controlled EV charging optimization at the distribution level to the aggregated EV charging demand profiles optimized for transmission system planning.
To this end, we distinguish between three cases of EV charging. 
In the first case, we determine the optimal charging demand results from the perspective of a centralized planner at the transmission level and assess their feasibility at the distribution level based on a top-down disaggregation model. 
The top-down model takes the perspective of a distribution grid operator who follows the centralized planner's setpoint, while explicitly accounting for distribution grid constraints and individual driving requirements.
In the second case, we optimize EV charging flexibility at the distribution level using a bottom-up approach.
The bottom-up approach takes the perspective of a distribution grid operator who minimizes costs, while subject to the electricity price signals from the central planner.
Finally, the outcomes of the top-down and bottom-up approaches are compared against an uncontrolled EV charging profile, which constitutes the third case.
The operational impacts of the three charging approaches are assessed in terms of their effects on charging profiles and grid violations at the distribution level, whereas charging costs and system costs are not within the scope of this work. 

With this framework, we address two research questions: 
(i) Can EV charging optimized from a centralized transmission system perspective be accommodated in distribution grid operations?
(ii) How does a distribution-centered optimization deviate from the centrally optimized EV charging profile?
The contributions of this work are threefold.
We demonstrate the impacts of centrally optimized EV charging dispatch on decentralized grid operations of integrated medium- and low-voltage grids. 
We provide a systematic comparison of a top-down transmission-centered EV charging approach with an uncontrolled and a bottom-up distribution-centered EV charging approach. 
Finally, we assess whether the impacts of the uncontrolled, top-down, and bottom-up EV charging approaches are robust to variations in the grid. To this end, we conduct a sensitivity analysis in which we vary PV deployment level, heat pump (HP) adoption rate, grid capacity, and simulation day.
This study is relevant to both power system researchers and grid operators, as it bridges the gap between transmission-level planning and distribution-level operations when dealing with distributed, flexible resources, such as EVs. 

The remainder of this paper is organized as follows. 
Section~\ref{sec:method} presents the methodological framework, introducing both the bottom-up and top-down approaches. 
Section~\ref{sec:cstudy} details the case study setup and describes the input data. 
Section~\ref{sec:results} reports the results, which are discussed in Section~\ref{sec:discussion_conclusion}. 

\section{Method} 
\label{sec:method}

\begin{figure}[b]
\includegraphics[width=\columnwidth]{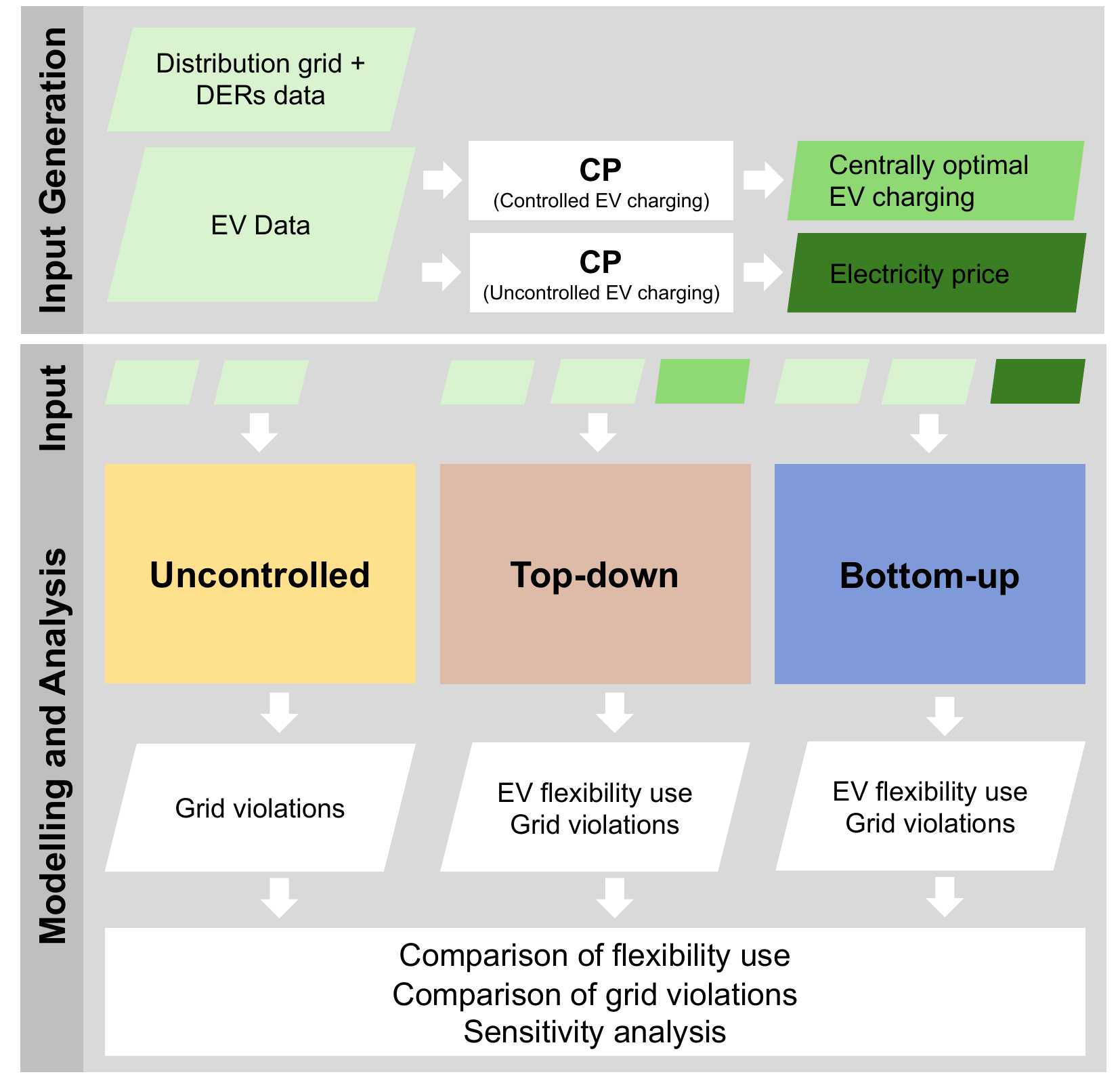}
\caption{Graphical representation of the methodology used to compare the uncontrolled, top-down, and bottom-up approaches for EV controlled charging.}
\label{fig:methodology}
\end{figure}

  
We assess the impacts of controlled and uncontrolled EV charging on the integrated MV and LV distribution grid. The methodological framework encompassing two controlled EV charging approaches: a top-down, transmission-level (centralized) approach and a bottom-up, distribution-level (decentralized) approach. We compare these two approaches with uncontrolled EV charging.
The uncontrolled charging approach is based solely on individual mobility needs and comfort preferences. Charging begins immediately after parking and proceeds at the maximum available power at the charging location. 
The top-down approach optimizes EV charging demand via a centralized generation expansion and operation planning (CP) model and disaggregates it to the distribution grid, while explicitly accounting for distribution grid constraints and individual driving requirements.
The bottom-up approach optimizes EV charging demand to minimize energy supply costs while respecting the distribution grid's operational limits. 
We analyze how each controlled charging approach exploits EV flexibility and benchmark both against the uncontrolled baseline. For each approach, we further assess its effects on grid performance, namely, voltage magnitudes and branch loading, under varying levels of DERs penetration and grid capacity. Voltage levels are considered acceptable when kept between $0.90$~$p.u.$~-~$1.10$~$p.u.$ of the nominal value, while thermal limits are satisfied when branch loading remains below $1.00$~$p.u.$ of rated capacity.
The overall methodology is illustrated in Figure~\ref{fig:methodology}, and detailed in Sections (\ref{sec:EV_modeling}–\ref{sec:bottomup}).

\subsection{EV modeling}
\label{sec:EV_modeling}
The EV charging model follows the development in~\cite{ParajelesSEST2024, Parajeles2025}. It provides geo-referenced data on driving, parking, charging behavior, and available flexibility at the individual vehicle level. 
The model produces a baseline charging schedule (i.e., uncontrolled charging where users decide to charge solely based on their mobility needs and EV's battery comfort range), and values for two flexibility metrics: (i) upper and lower flexible power bounds, which define the maximum and minimum charging power at any time step calculated based only on available idle parking time in each charging event), and (ii) a maximum flexible charging energy metric, which constrains total allowed daily energy deviations from the baseline charging profile. This quantification method captures the flexibility available when users’ charging decisions, comfort range preferences, and driving needs are respected, thereby representing the achievable flexibility with minimal user disruption and, consequently, a high likelihood of participation.
The high granularity enables the mapping of the EV charging and flexibility to low-voltage (LV) nodes. 
Additionally, the model provides aggregated charging demand and flexibility data, rendering it also suitable for analyzing EV charging flexibility at lower geographical resolutions. 

\subsection{Centralized generation expansion and operation planning (CP) model}\label{sec:CP}


The centrally optimized EV charging demand is computed using CentIv, a centralized generation expansion planning and operation optimization model. CentIv is part of Nexus-e, an integrated energy systems modeling platform that includes models for decentralized generation expansion planning and operation, power system security, electricity markets, and macro-economic analysis~\cite{Gjorgiev2022}. 
CentIv optimizes investment and operational decisions over a one-year horizon with hourly resolution, while fully representing the transmission grid~\cite{Raycheva2021}. 

The EV charging demand and flexibility data described in Section~\ref{sec:EV_modeling} are aggregated and provided as inputs to CentIv at each node of the transmission network. Specifically, the model receives three inputs: (1) the aggregated baseline EV charging profile, (2) the upper and lower power bounds, and (3) the maximum daily flexible energy. 
Based on these inputs, CentIv centrally optimizes the temporal allocation of aggregated EV charging flexibility at each transmission node over an annual time horizon, subject to a weekly energy balance constraint that guarantees fulfillment of EV charging energy demand.

The resulting dispatch defines the centrally optimized EV charging demand, which serves as the input for subsequent distribution level disaggregation.

\subsection{Top-down controlled charging approach} \label{sec:topdown}

The top-down model translates the centrally optimized EV charging demand described in Section~\ref{sec:CP} to the distribution grid level, reflecting the perspective of a distribution grid operator who follows a setpoint provided by the CP. 
The distribution system operator in the top-down approach has two priorities: (i) Dispatch the EVs connected to its grid while adhering to the centrally optimized charging strategy, and (ii) comply with the distribution grid constraints.
The model takes as input the optimal EV charging dispatch at the transmission node supplying a selected distribution grid of interest, using its temporal shape but rescaling peak power and total energy to match the charging demand flexibility available at the considered distribution grid.


This is achieved in two steps. First, we normalize the centrally optimized EV charging dispatch using the upper power bound of EV flexibility at the corresponding transmission node, as this upper bound represents the maximum connected charging power. Second, the normalized dispatch is rescaled using the aggregated upper flexible power bound for the set of vehicles that charge within the considered distribution grid supply area, ensuring that the charging profile is realistic with respect to the number of vehicles connected to that distribution grid and their charging rates. The resulting normalized and rescaled charging profile is hereafter referred to as the \textit{CP} profile.

Finally, to allocate charging demand to the distribution grid nodes, we solve the optimization problem as given by~(\ref{eq:obj_centralized})~-~(\ref{eq:relaxed_const_centralized}). Its purpose is to reproduce the CP profile at the distribution level by scheduling each vehicle's charging within its supply area, while accounting for local grid and DER constraints. This formulation yields a distribution-grid-friendly charging configuration that prioritizes adherence to the CP profile. We refer to the charging profile resulting from this optimization as the \textit{top-down} profile.

\begin{align}
\min \ & \sum_{t\in\mathcal{T}} z_t 
         + \rho \sum_{t\in\mathcal{T}} s_t,
        && \rho \ll 1 \label{eq:obj_centralized}\\
\text{s.t.} \quad
& -z_t \le \sum_{v} p_{v,t} - \bar P_t \le z_t, 
   && \forall t \label{eq:diff_centralized} \\
& L_t \le \sum_v p_{v,t} \le U_t,
  && \forall t \label{eq:plimits_centralized} \\
& SOE_{v,t+1} = SOE_{v,t} + \eta_v p_{v,t}\Delta t, 
  && \forall v,t \label{eq:ebattery_centralized}\\
& SOE^{\min}_v \le SOE_{v,t} \le SOE^{\max}_v, 
  && \forall v,t \label{eq:blimits_centralized} \\
& SOE_{v,e} \ge E^{\text{trip}}_{v,e}, 
  && \forall v,e \in \mathcal{S}_v \label{eq:dlimits_centralized}\\
& \sum_v SOE_{v,|\mathcal{T}|} \ge SOE^{\text{unctrl,end}} \label{eq:soc_centralized} \\
& 0 \le G^{\mathrm{PV}}_t \le \bar G^{\mathrm{PV}}_t, && \forall t \label{eq:pv_bounds}\\
& c^{\mathrm{PV}}_t = \bar G^{\mathrm{PV}}_t - G^{\mathrm{PV}}_t, && \forall t \label{eq:pv_curt_def}\\
& \sum_{t \in \mathcal{T}} c^{\mathrm{PV}}_{t}\;\le\;\bar C^{\mathrm{PV}} \label{eq:pv_centralized}\\
& (P_t,Q_t,V_t) \in \mathcal{F}_{\mathrm{LinDistFlow}},
  && \forall t \label{eq:df_centralized}\\
& g(P_t,Q_t,V_t) \le s_t,\quad s_t \ge 0,
  && \forall t \label{eq:relaxed_const_centralized}
\end{align}

The optimization's objective function in~(\ref{eq:obj_centralized}) and the constraint in~(\ref{eq:diff_centralized}) minimize the difference between the CP and the top-down profile at each time step. 

Additionally,~(\ref{eq:plimits_centralized})–(\ref{eq:soc_centralized}) model strict constraints on vehicle charging and driving needs, defined by the full-resolution EV use, charging, and flexibility data detailed in Section~\ref{sec:EV_modeling}. The optimization also models the operation of rooftop PV generation (\ref{eq:pv_bounds}) and curtailment (\ref{eq:pv_curt_def})-(\ref{eq:pv_centralized}). PV maximum curtailment rates are consistent with those obtained from the CP optimization.

Finally, power flows are modeled with the linearized Dist-Flow formulation in~(\ref{eq:df_centralized}) \cite{BaranWu1989}. 
In this modeling approach, grid operational limits on branch loading and voltage magnitude (\ref{eq:relaxed_const_centralized}) are modeled as soft constraints with the corresponding slack variable weighted by $\rho \ll 1$ in the objective function.

    
\subsection{Bottom-up controlled charging approach}\label{sec:bottomup}

The bottom-up model represents an optimal EV charging behavior from the perspective of a cost-minimizing distribution grid operator. 
In practice, EV charging control would typically be carried out by an EV aggregator, whose objective is to dispatch EV demand in a cost-optimal manner. 
Such an aggregator may lack detailed knowledge of distribution grid constraints or deprioritize the interests of the DSO. 
Therefore, we emphasize that the bottom-up approach is used to represent the perspective of a distribution system operator.
The distribution system operator in the bottom-up approach has two priorities: (i) minimize the net cost of electricity exchanged with the transmission level, and (ii) comply with the distribution grid constraints.
To represent the first priority, i.e., cost minimization, we use wholesale electricity prices derived from a nationwide simulation assuming uncontrolled charging conditions (i.e., no charging flexibility) using the centralized generation and expansion planning model described in Section~\ref{sec:CP}.

The charging behavior of each vehicle is determined by solving the optimization problem defined by~(\ref{eq:obj_bottomup})~-~(\ref{eq:relaxed_const_bottomup}). 


\begin{align}
\min \ & \sum_{t\in\mathcal{T}} c_{el,t}\cdot (D_{\mathrm{res},t} - G_{\mathrm{res},t}) 
      + \beta\sum_{t} \big(
          \textbf{s}_t
        \big), && \beta \gg 1 \label{eq:obj_bottomup} \\[2pt]
\text{s.t.} \quad
& L_t \le \sum_v p_{v,t} \le U_t,
  && \forall t \label{eq:plimits_bottomup} \\
& SOE_{v,t+1} = SOE_{v,t} + \eta_v p_{v,t}\Delta t, 
  && \forall v,t \label{eq:ebattery_bottomup}\\
& SOE^{\min}_v \le SOE_{v,t} \le SOE^{\max}_v, 
  && \forall v,t \label{eq:blimits_bottomup} \\
& SOE_{v,e} \ge E^{\text{trip}}_{v,e}, 
  && \forall v,e \in \mathcal{S}_v \label{eq:dlimits_bottomup}\\
& \sum_v SOE_{v,|\mathcal{T}|} \ge SOE^{\text{unctrl,end}} \label{eq:soc_bottomup} \\
& 0 \le G^{\mathrm{PV}}_t \le \bar G^{\mathrm{PV}}_t, && \forall t \label{eq:pv_bottomup}\\
& c^{\mathrm{PV}}_t = \bar G^{\mathrm{PV}}_t - G^{\mathrm{PV}}_t, && \forall t \label{eq:pv_curt_def_bottomup}\\
& \sum_{t \in \mathcal{T}} c^{\mathrm{PV}}_{t}\;\le\;\bar C^{\mathrm{PV}} \label{eq:pv_curt_bottomup}\\
& (P_t,Q_t,V_t) \in \mathcal{F}_{\mathrm{LinDistFlow}},
  && \forall t \label{eq:df_bottomup}\\
& g(P_t,Q_t,V_t) \le s_t,\quad s_t \ge 0,
  && \forall t \label{eq:relaxed_const_bottomup}
\end{align}

The first term in the objective function minimizes the cost of electricity purchases (\ref{eq:obj_bottomup}). Similarly to the top-down approach, vehicle charging is subject to strict constraints that ensure compliance with driving and charging requirements~(constraints (\ref{eq:plimits_bottomup})–(\ref{eq:soc_bottomup})), as well as constraints on the operation of rooftop PV systems, i.e.~(\ref{eq:pv_bottomup})-(\ref{eq:pv_curt_bottomup}).

To represent compliance with grid restrictions, the problem includes soft constraints, implemented through a large penalty factor ($\beta \gg 1$) in the objective function, relaxing grid operational limits on branch loading and voltage magnitude (\ref{eq:relaxed_const_bottomup}) determined by the linear DistFlow formulation (\ref{eq:df_bottomup}) \cite{BaranWu1989}. The large penalty reflects the operator’s strong incentive to avoid violating these grid constraints.
With this bottom-up modeling approach, we model an EV charging behavior aligned with a distribution operator's priorities. 
We refer to the charging profile resulting from~(\ref{eq:obj_bottomup})~-~(\ref{eq:relaxed_const_bottomup}) as the \textit{bottom-up} profile.

\section{Case study and data} 
\label{sec:cstudy}



The analysis focuses on Switzerland in 2050, in accordance with the ambitious timeline for full decarbonization. Switzerland’s federal energy strategy targets \SI{45}{TWh} renewable generation in addition to its existing hydro capacity~\cite{swissconfederation2023}, accompanied by a progressive phase-out of nuclear power in favor of solar photovoltaics. According to the national energy perspectives~\cite{Energieperspectives2019}, most of this PV capacity is expected to be deployed on rooftops, reaching around \SI{34}{TWh} of annual generation.

Switzerland will thus see a strong shift toward distributed renewable production, increasing the relevance of the interaction between centralized system planning and decentralized operation. In parallel, the decarbonization strategy foresees substantial electrification of the heating and transport sectors~\cite{Energieperspectives2019}. 
As a result, the 2050 Swiss power system is expected to face increased electricity demand, along with high shares of non-dispatchable renewables, underscoring the need for flexible and coordinated operation of distributed energy resources.


\subsection{EV charging demand and flexibility data}\label{subsec:cs_ev_df}




EV usage data, charging demand, and flexibility are adopted from~\cite{ParajelesSEST2024, Parajeles2025}. In our 2050 scenario, we assume full electrification of all privately-owned passenger vehicles that can feasibly be replaced by EVs. This assumption represents an upper bound on EV flexibility's potential impact - a useful reference point given that the quantification model is designed to minimize user disruption.

For the CP model, we utilize the complete EV dataset generated by~\cite{ParajelesSEST2024, Parajeles2025}, which includes approximately $3.2$~million vehicles. Vehicle-level information is aggregated for each of the $129$ transmission supply areas in Switzerland. The national annual demand results in about $\SI{12}{TWh}$ and a weekday peak demand of around $\SI{3.07}{\giga\watt}$. Around $68\%$ of the daily charged energy is quantified as flexible during weekdays, with an average upward flexibility of $\SI{1.2}{\giga\watt}$ and downward flexibility of about $\SI{500}{\mega\watt}$.

Electrified vehicles are assumed to operate with battery capacities between $\SI{70}{kWh}$ and $\SI{120}{kWh}$. This range is informed by Swiss EV registration data~\cite{AutoSchweiz}, which reports an average battery size of about $\SI{70}{kWh}$ to date, and by ongoing market trends toward larger capacities. Charging rates vary by location: $\SI{7}{kW}$ at home, $\SI{11}{kW}$ at workplaces, and $\SI{22}{kW}$ at public stations, consistent with projections for Switzerland~\cite{Gschwendtner2023, Ramsebner2023}.

For both the top-down and bottom-up models, we consider only vehicles with charging events located within the supply area of the selected distribution grid under study (see Section~\ref{subsec:cs_bottom_up}). This corresponds to approximately $17,442$ vehicles, for which geographic-specific charging locations are mapped to the nearest distribution grid node in the LV grids. For the selected grid, we find an annual charging demand of \SI{25}{GWh} and a peak charging demand of \SI{8.4}{MW}. Within this area, approximately $55~\%$ of daily charged energy is quantified as flexible during weekdays, with an average upward flexibility of \SI{3.0}{MW} and downward flexibility of about \SI{1.3}{MW}.

\subsection{Setup for the CP simulation}\label{subsec:cs_top_down}
The CP model computes the centrally optimized EV charging demand by performing generation expansion planning and hourly dispatch within Switzerland, considering input data on candidate generation capacity, renewable resource availability time series, and cost parameters, as detailed in~\cite{ParajelesCentralized2025, Raycheva2025}.
The Swiss 2050 electricity demand is based on the sum of conventional electricity demand from the Swiss Energy Perspectives 2050+~\cite{Energieperspectives2019}, EV charging demand for passenger cars from~\cite{Parajeles2025}, and HP demand from~\cite{Guo2026}. Electricity demand in neighboring countries is based on the ENTSO-E Ten-Year Network Development Plan (TYNDP) 2022 data~\cite{tyndp2022}.
Furthermore, all national renewable energy and rooftop PV targets for 2050 are enforced~\cite{Energieperspectives2019}, and nuclear generation is assumed to be fully phased out in Switzerland, in line with the current Swiss energy policy~\cite{swissconfederation2023}.

\begin{figure}[h]
\centering
\includegraphics[width=1\columnwidth]{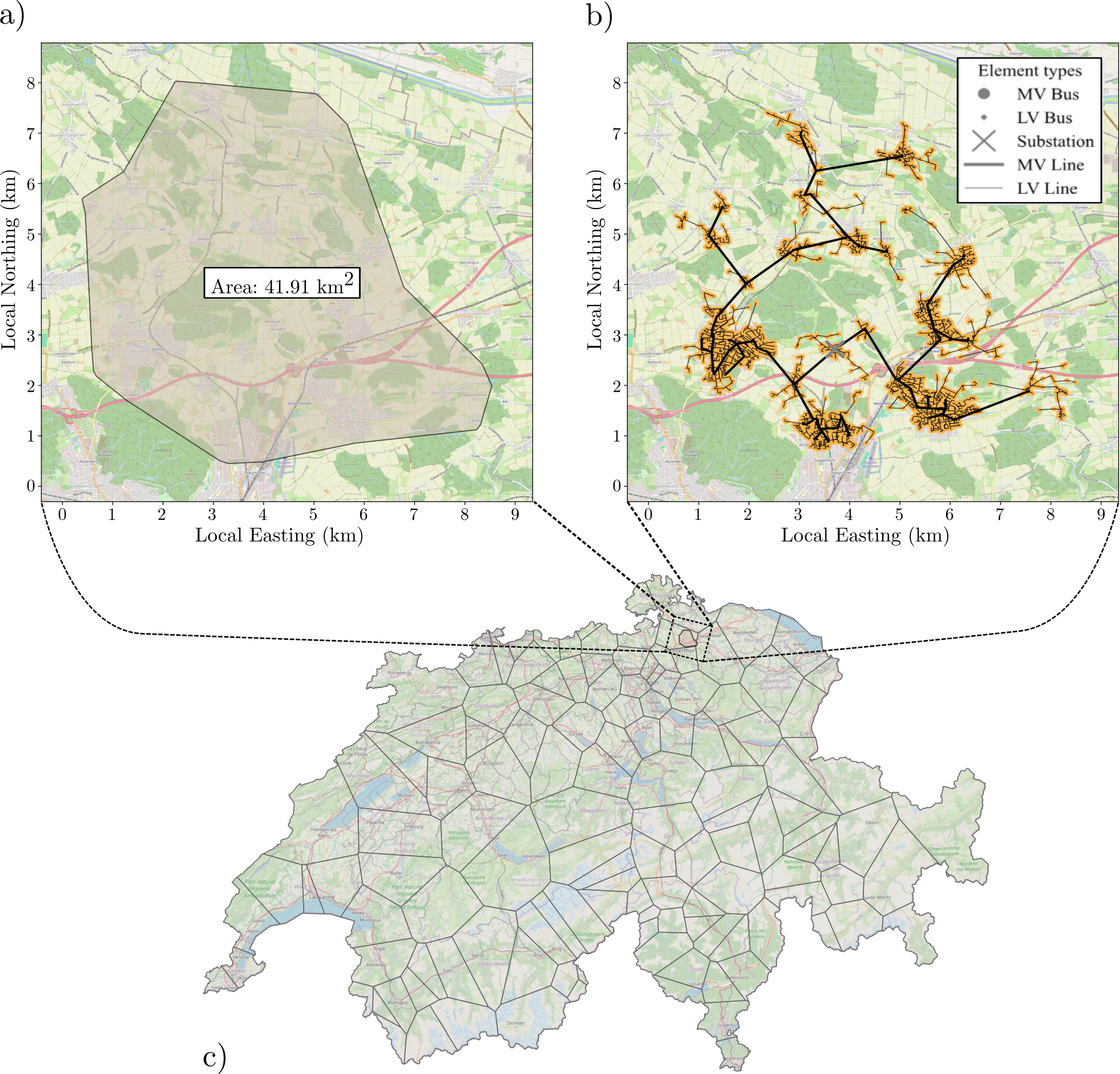}
\caption{Graphical representation of a) the distribution grid supply area, b) the topology of the distribution grid, and c) the location of the selected distribution grid in the national-scale transmission supply areas.}
\label{fig:grid_location}
\end{figure}

Although the CP model simulates the entire Swiss 2050 transmission system, we focus on results for the transmission node supplying selected medium- and low-voltage distribution grids, which are later analyzed using the top-down and bottom-up models introduced in Section~\ref{sec:method}. Figure~\ref{fig:grid_location} illustrates the 129 transmission supply areas in Switzerland, along with a close-up of the specific supply area corresponding to the selected distribution grid including the grid's topology (see Section~\ref{subsec:cs_bottom_up} for details).


\subsection{Setup for the distribution grid simulation}\label{subsec:cs_bottom_up}
%

The distribution grid is considered in both the top-down and the bottom-up models. As shown in Figure~\ref{fig:grid_location}, we select a grid located in the northeast of Switzerland, covering predominantly urban and peri-urban areas. The grid supplies an area of approximately $\SI{40}{km^2}$ in the vicinity of the city of Winterthur.
The expected high concentration of electric vehicles and electrified heating loads associated with the urbanization level of this area in 2050 makes this a relevant case study.  

The complete grid topology is obtained from the repository of synthetic Swiss grids presented in~\cite{Oneto2025}. We model both the medium-voltage (MV) ($\SI{20}{kV}$) and LV ($\SI{0.4}{kV}$) levels as fully radial structures. The considered grid has 109 MV buses, 108 MV lines, 54 transformers, 8105 LV buses, and 8051 LV lines. The dataset from~\cite{Oneto2025} provides peak power information for 2050, but does not include time-series load profiles. To reconstruct temporal demand patterns, we use data from~\cite{opendataswiss2025}, which contains a historical repository of aggregated conventional demand for the Winterthur region.
We assume that the shape of conventional demand profiles remains constant up to 2050. Each consumer node in the grid is assigned a normalized daily load profile. These profiles are randomly sampled from a window spanning 10 days before and after the representative winter and summer days considered in the study and defined in Section~\ref{subsec:cs_sensitivity}. The sampled profiles are then scaled according to each node’s peak demand. This procedure avoids artificial coincident peak loads across the grid while maintaining realistic spatial and temporal diversity of consumption.




For the PV capacity installed in the distribution grid, we use results from the national-scale analysis conducted in the Nexus-e platform (specifically, CentIv combined with the decentralized investment module), which provides municipality-level PV investment and corresponding capacities. The studied distribution grid supplies nine municipalities (some only partially) in the northeast of Switzerland. The total PV capacity connected to this specific distribution grid is approximately \SI{460}{MW}.

To allocate this capacity spatially, rooftops are sampled from the pool of suitable buildings within the grid area~\cite{SFOS2025}. We leverage data from the Sonnendach dataset~\cite{opendataswiss2023}, which provides per-building information, including suitability, average irradiation, surface area, and optimal tilt of the associated PV system. The peak power rating of each sampled rooftop is computed assuming a panel efficiency of $25\%$ under standard test conditions and $90\%$ rooftop surface utilization. The sampling continues until the target capacity is reached. Generation time series are created for the considered simulated summer and winter days, using irradiation data from~\cite{opendataswiss2023}.



By 2050, it is projected that approximately $55\%$ of the heating demand in the Swiss building sector will be met by heat pumps~\cite{SwissFederalOfficeofEnergy2025}. Heat demand profiles are derived following the method described in~\cite{Zapparoli2025}. For each building within the supply area, a heat pump sizing procedure is first performed based on the characteristics of the building obtained from~\cite{SFOS2025}. Key parameters include the energy reference area, base surface area, building thermal conductivity and capacitance, and the outdoor temperature profile corresponding to the building’s location~\cite{Zapparoli2025}.

Consistent with national electrification targets, we randomly allocate heat pumps to $50\%$ of the buildings in each of the categories defined by~\cite{SFOS2025} (i.e., residential, mixed-use, and commercial buildings), restricting the selection to units with rated capacities below $\SI{100}{kW}$. We then generate time series of heat demand and corresponding electricity consumption using the approach in~\cite{Zapparoli2025}, assuming a target indoor temperature of $\SI{20}{\degreeCelsius}$ and outdoor temperature data for the representative winter and summer days. Cooling demand is not considered, as it is negligible under Switzerland's climatic conditions and current building context. 

\subsection{Study setup and sensitivity analysis}\label{subsec:cs_sensitivity}

The top-down and bottom-up optimization models are large-scale second-order cone programming problems comprising more than 10 million decision variables, primarily due to the explicit representation of a large vehicle fleet with high spatial resolution. 
Executed on 48 parallel threads, each optimization problem requires approximately 10~h. Preserving the high spatial resolution of the analysis, therefore, imposes practical limits on the temporal scope of the study and restricts simulations over extended time horizons.

To balance computational tractability with temporal representativeness, the optimizations are solved over 24-hour time horizons, focusing on selected days in two extreme seasons: winter and summer. Representative days are identified for January and July, corresponding to the central months of each season and capturing substantially different grid operating conditions resulting from variations in heating demand, photovoltaic generation, and overall electricity consumption.

Within each season, the representative day (D) is selected by examining the distributions of electricity demand and PV generation and choosing a day that falls within the 40th--60th percentile range for both indicators. This criterion aims to avoid atypical operating conditions and instead capture days that are broadly representative of seasonal system behavior. To reduce sensitivity to the exact temporal selection, the analysis window is extended to include the day immediately preceding (D-1) and following (D+1) each representative day D. Consequently, all scenarios (i.e., main study and sensitivity analyses) are evaluated over three days for both winter and summer conditions.

This temporal setup enables a direct comparison of centralized (top-down) and decentralized (bottom-up) charging control approaches and their ability to exploit EV charging flexibility under distinct seasonal operating conditions while improving the robustness of the resulting observations.

To further isolate the influence of individual system characteristics on the performance of the charging approaches, we perform one-way sensitivity analyses in which each parameter is varied independently, one at a time, while all other inputs are held at their reference values. Specifically, we consider variations in rooftop PV and HP deployment, implemented by reducing the number of deployed systems by 25\% and 50\%, as well as variations in grid capacity, increased by 25\% and 50\% relative to the reference scenario. Changes in grid capacity are implemented by increasing network transfer capability through proportional reductions in branch impedance values and the corresponding increase in line capacities.   
This setup enables clear attribution of observed effects to isolated changes in PV, HP, or grid capacity, and supports a robust comparison of charging approaches across seasonal, daily, and grid-related variations.

\section{Results} 
\label{sec:results}

In this section, we compare the performance of the top-down and bottom-up controlled charging approaches with that of the uncontrolled charging approach. 
First, we analyze the resulting daily EV charging profiles for the simulated winter and summer days.
Next, we present the results on grid violations and profile deviations.
Additionally, we present the outcomes of the sensitivity analysis conducted with respect to deployment levels of rooftop PV systems, HPs, and grid capacity. 
Finally, we provide a geographical representation of the grid violations, linking their spatial distribution to local patterns of PV generation and HP demand. 

\subsection{Daily profiles}
\label{sec:results_daily_profiles}
\begin{figure*}
  \centering
    \includegraphics[width=\textwidth]{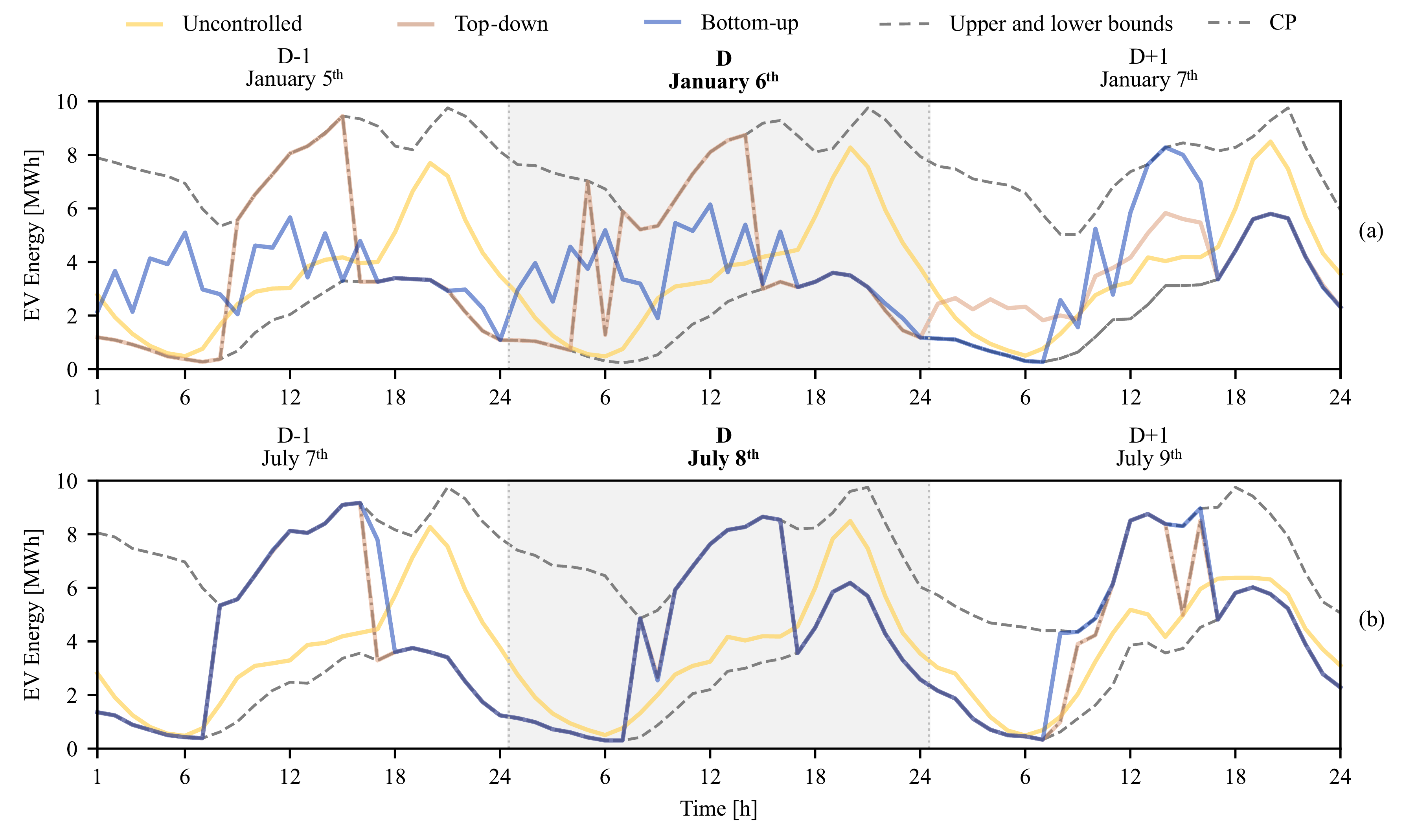}
  \caption{Electric vehicle consumption profiles for the simulated winter day and summer days: the representative days (D), and the days before (D-1) and after (D+1). 
  Each panel reports the uncontrolled, the top-down and bottom-up controlled charging cases, the aggregated electric vehicle flexibility bounds, and the reference profile provided by the CP model.}
  \label{fig:profiles}
\end{figure*}

Figure~\ref{fig:profiles} shows the daily profiles for the different EV charging approaches described in Section~\ref{sec:method}, for the simulated winter and summer days.
Overall, controlled charging approaches shift the uncontrolled EV evening peak towards noon or, more generally, to daytime hours.  

In both seasons, the top-down approach (which models distribution grid constraints) often exhibits nearly identical dispatch patterns and reproduces the CP profile with high accuracy.
This result is non-trivial, as it demonstrates that the CP model, based on aggregated EV data and agnostic to the distribution grid constraints, can identify an optimal charging profile compatible with individual vehicle driving patterns.
Moreover, the top-down profile follows closely the EV flexibility bounds for most of the day. 
This indicates that the centralized optimization, agnostic to distribution grid constraints, fully exploits the available EV charging flexibility to identify the cost-optimal transmission grid dispatch.
Figure~\ref{fig:profiles}, however, also reveals exceptions to this ideal behavior.
In the last simulated winter day (D+1), the top-down charging profile deviates from the CP profile, to ensure that the daily final SoE of the top-down charging approach is at least equal to that of the uncontrolled charging approach (constraint~\ref{eq:soc_centralized} in Section~\ref{sec:topdown}).
This requirement in the top-down model is introduced because of its limited 24-hour optimization horizon.
By contrast, the CP model is optimized over a longer time horizon and, therefore, does not impose a daily final SoE target, relying instead on weekly energy balancing.
Consequently, discrepancies between the CP and top-down charging profiles may arise from differences in temporal energy constraints rather than from limitations in charging flexibility or grid operability.

The bottom-up approach, on the other hand, produces notably different results for winter and summer.
In winter, the bottom-up profile differs significantly from the top-down profile.
In fact, the bottom-up approach gives a higher priority to relieving distribution grid congestions. 
Therefore, this approach reduces grid congestions by lowering peak demand and distributing EV charging more evenly throughout the day, while still respecting the vehicle constraints.
Although the temporal patterns of the bottom-up profile differ from those of the top-down profile, they present similar characteristics: both concentrate charging demand during daytime hours and reduce the evening peak. 
In contrast, during summer, the bottom-up and top-down profiles nearly overlap. 
The summer grid context differs significantly from that of winter, as PV generation increases while HP demand is at a minimum. 
Consequently, enabling grid-friendly PV injections becomes a common objective for both centralized and decentralized perspectives. 

\subsection{Grid violations and charging profiles}
\begin{figure}
\centering
\includegraphics[width=1\linewidth]{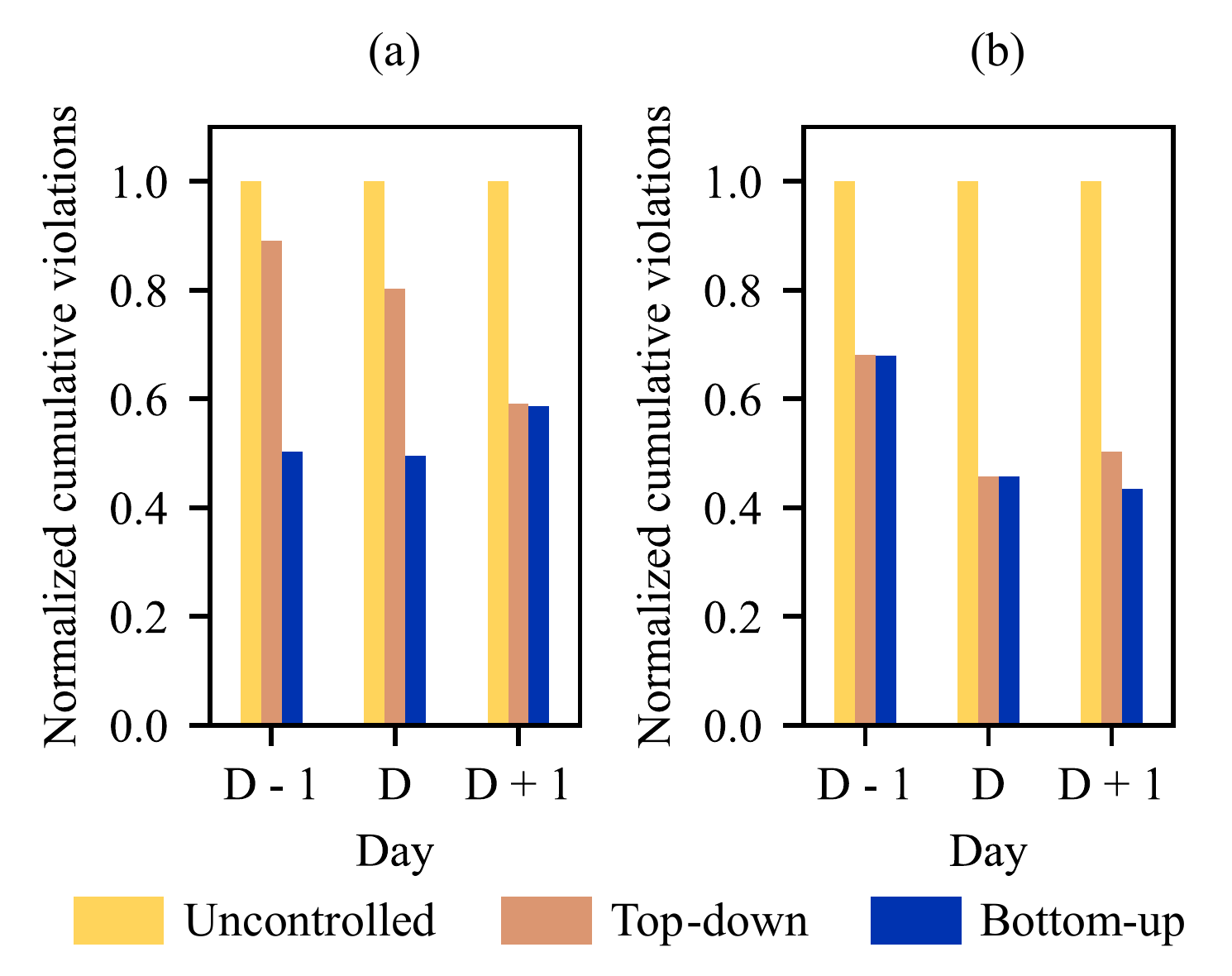}
\caption{Grid violations for the simulated days. The three EV charging approaches (uncontrolled, top-down, and bottom-up) are compared for the winter days (panel (a)) and the summer days (panel (b)).}
\label{fig:sensitivity_day_violations}
\end{figure}

Figure~\ref{fig:sensitivity_day_violations} shows the grid violations observed with the uncontrolled, top-down, and bottom-up charging approaches for the simulated winter (panel (a)) and summer (panel (b)) days.
The cumulative violations are computed as the sum of the per-unit (p.u.) violations of voltage and branch loading limits observed throughout the day with a simulation resolution of 1 hour. 
These values are then normalized with respect to the day's uncontrolled EV charging results.

The figures show that the top-down profile consistently reduces grid violations in both seasons by effectively dispatching available charging flexibility, although it is based on a CP model's centrally optimized EV charging demand, which is agnostic to distribution grid information.
These results reflect the emerging paradigm of future power systems, characterized by increasingly distributed generation and heterogeneous, flexible demand, which together facilitate stronger alignment between central and local operational objectives.
The bottom-up approach can further reduce violations by identifying more grid-friendly charging schedules.
However, this improvement is observed mostly in the winter days.
In the summer days, flexibility is dispatched similarly to the CP model, as shown in Figure~\ref{fig:profiles}. 
In summer, PV injection–induced violations become the dominant operational constraint.
Under these conditions, both the top-down and bottom-up approaches are driven toward the same narrow window of grid-friendly charging opportunities, leaving little room for the bottom-up method to differentiate its schedule from the top-down profile.

\begin{figure}
\centering
\includegraphics[width=1\linewidth]{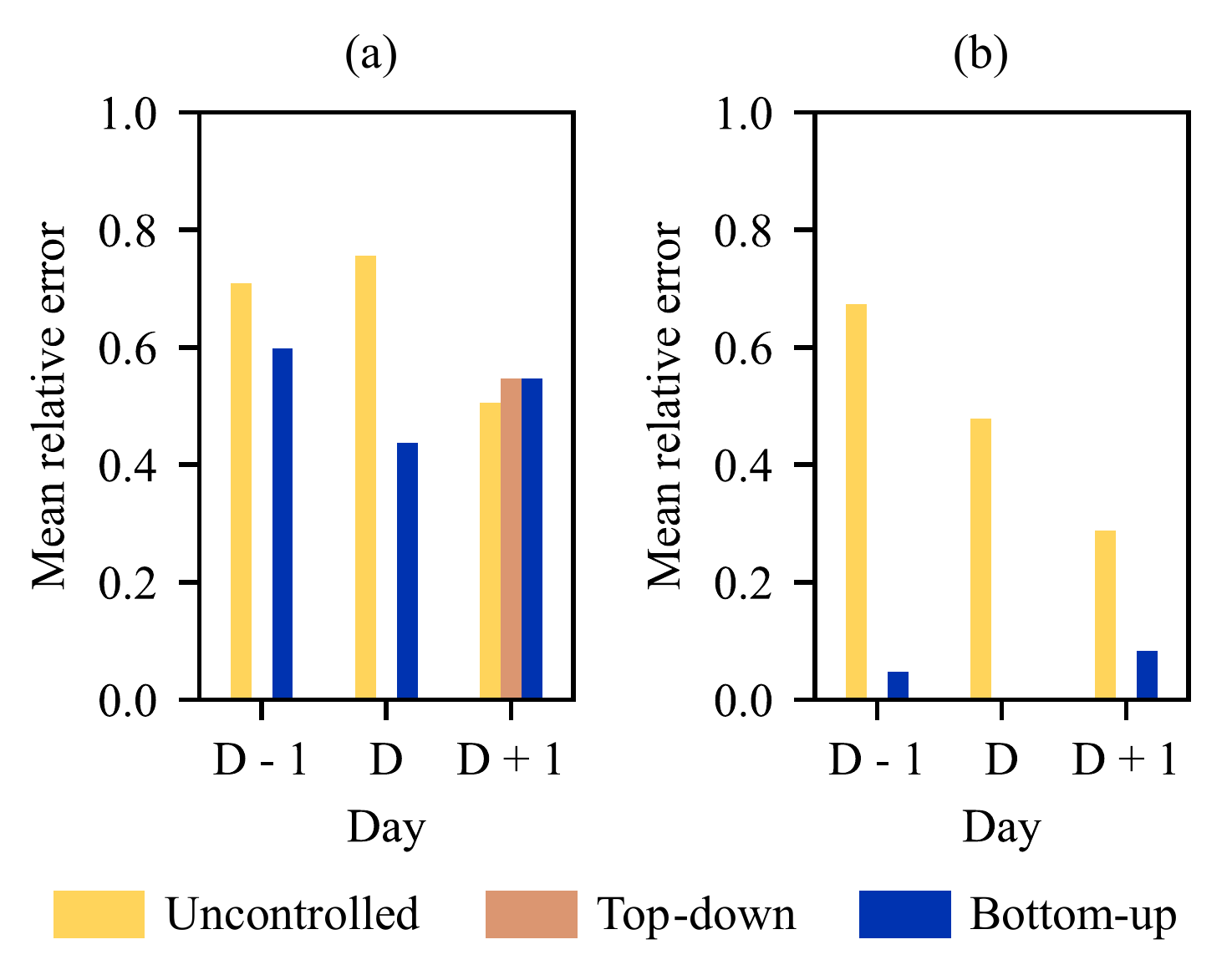}
\caption{Mean relative deviations of the EV charging approaches from the CP profile, for the simulated days. The three EV charging approaches (uncontrolled, top-down, and bottom-up) are compared for the winter days (panel (a)) and the summer days (panels (b)).}
\label{fig:sensitivity_day_profile}
\end{figure}

Figure~\ref{fig:sensitivity_day_profile} illustrates the deviation of the uncontrolled, top-down, and bottom-up profiles from the CP profile.
This deviation is quantified using the mean relative error, computed over the full daily profiles for the simulated winter (panel (a)) and summer (panel (b)) days.

The figure shows that the top-down approach typically reproduces the CP profile with high accuracy, regardless of the season. 
Additionally, the figure shows that the reduced grid violations of the bottom-up approach come at the expense of a deviation from the CP profile, suggesting that EVs would need to be charged differently throughout the simulated days.
Nevertheless, the bottom-up charging profiles remain significantly closer to the CP profile than the uncontrolled one.

\subsection{Sensitivity analysis}
\begin{figure*}
\centering
\includegraphics[width=\textwidth]{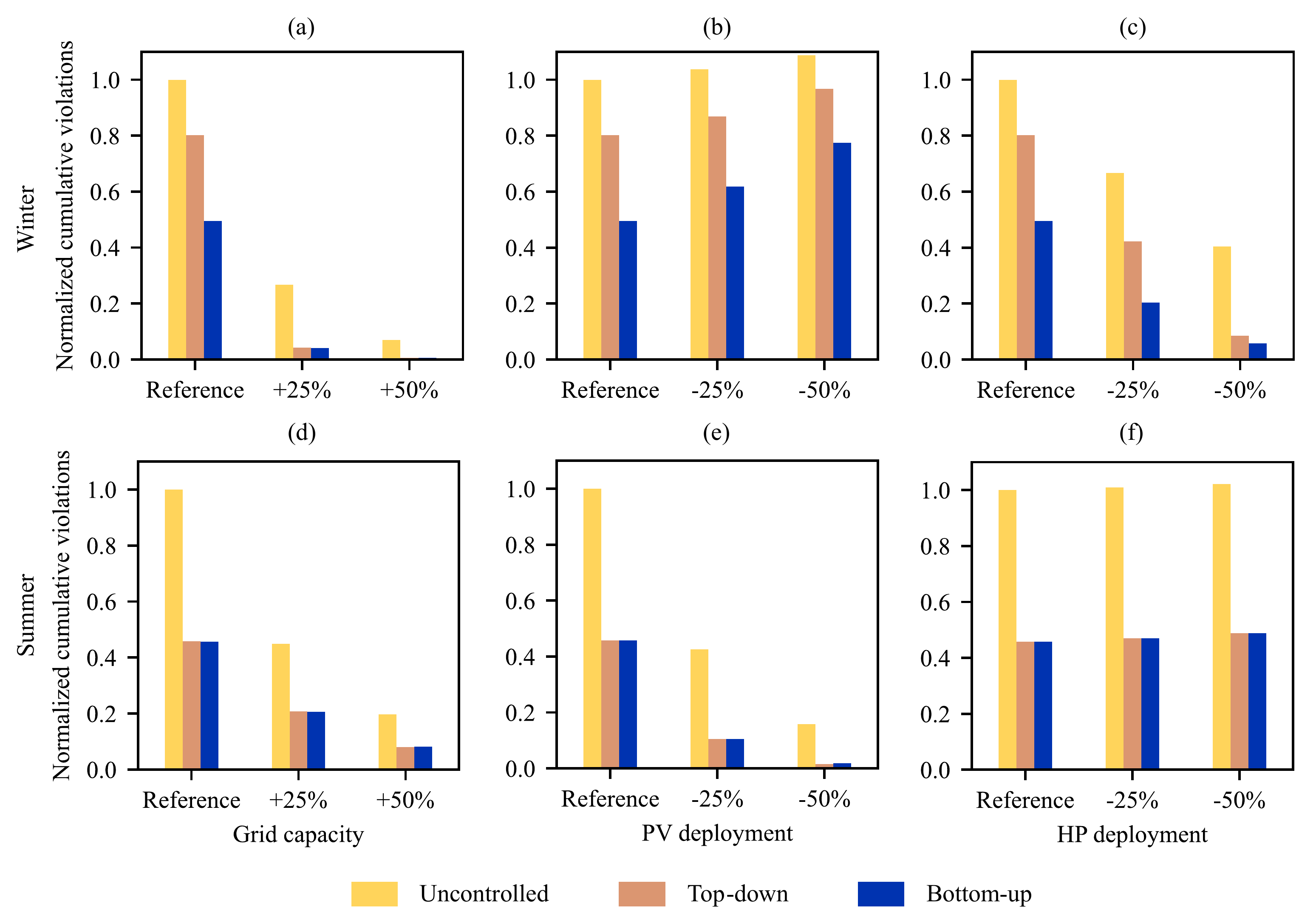}
\caption{Sensitivity analysis results for grid violations under different experimental settings. The three EV charging approaches (uncontrolled, top-down, and bottom-up) are compared for both the representative winter day (panels (a)--(c)) and the representative summer day (panels (d)--(f)). Panels (a) and (d) show the impact of different levels of grid reinforcement, panels (b) and (e) the impact of PV deployment, and panels (c) and (f) the impact of HP deployment.}
\label{fig:sensitivity_violations}
\end{figure*}

Figure~\ref{fig:sensitivity_violations} shows the grid violations observed with the uncontrolled, top-down, and bottom-up charging approaches for the winter (panels (a)--(c)) and summer (panels (d)-(f)) cases. The panels compare the \textit{Reference} scenario with the results obtained from varying grid capacity (panels (a) and (d)), PV deployment (panels (b) and (e)), and HP deployment (panels (c) and (f)).
The \textit{Reference} scenario is defined as the representative day scenario, with standard values for grid capacity, PV, and HP deployment.
The cumulative violations in Figure~\ref{fig:sensitivity_violations} are normalized with respect to the \textit{Reference} scenario with uncontrolled EV charging.

The sensitivity analysis cases on grid capacity, PV deployment, and HP adoption demonstrate that the results are robust to different underlying grid conditions. 
It further identifies that grid capacity has a major influence on reducing grid violations.
However, in these results, a grid capacity increase does not solve all violations, as the capacity increase considered here corresponds to a uniform upscaling of all grid components rather than a targeted or optimized reinforcement strategy.
Additionally, the sensitivity analysis on grid capacity shows that, even with 25\% and 50\% grid reinforcement, substantial violations persist under the uncontrolled EV charging approach. 
At the same time, this is not the case for the controlled charging approaches.
On a different note, a grid capacity increase is an impractical solution to grid violations, due to its high investment costs.

The reduction in PV and HP deployment has contrasting effects on the winter and summer days.
In summer, a PV deployment reduction leads to significant improvements, indicating that excessive PV generation causes additional stress on the grid.
This suggests the need to curtail excessive PV injections through PV control schemes that limit generation when necessary.
In winter, by contrast, lower PV deployment results in more grid violations.
This effect is primarily driven by the high electricity demand from heat pumps during this season, which can be partially supplied by local PV generation, thereby mitigating potential grid stress.  
These findings suggest that the distribution grid benefits from distributed PV generation when it supports local self-consumption.
Similarly, reduced HP deployment has beneficial effects in winter, as it lowers total demand and helps maintain a better balance between local supply and demand.
Conversely, in summer, HP reduction has little to no impact on grid violations, since these are primarily driven by excessive PV generation rather than high demand, and HP demand in summer is low.

\begin{figure*}
\centering
\includegraphics[width=\textwidth]{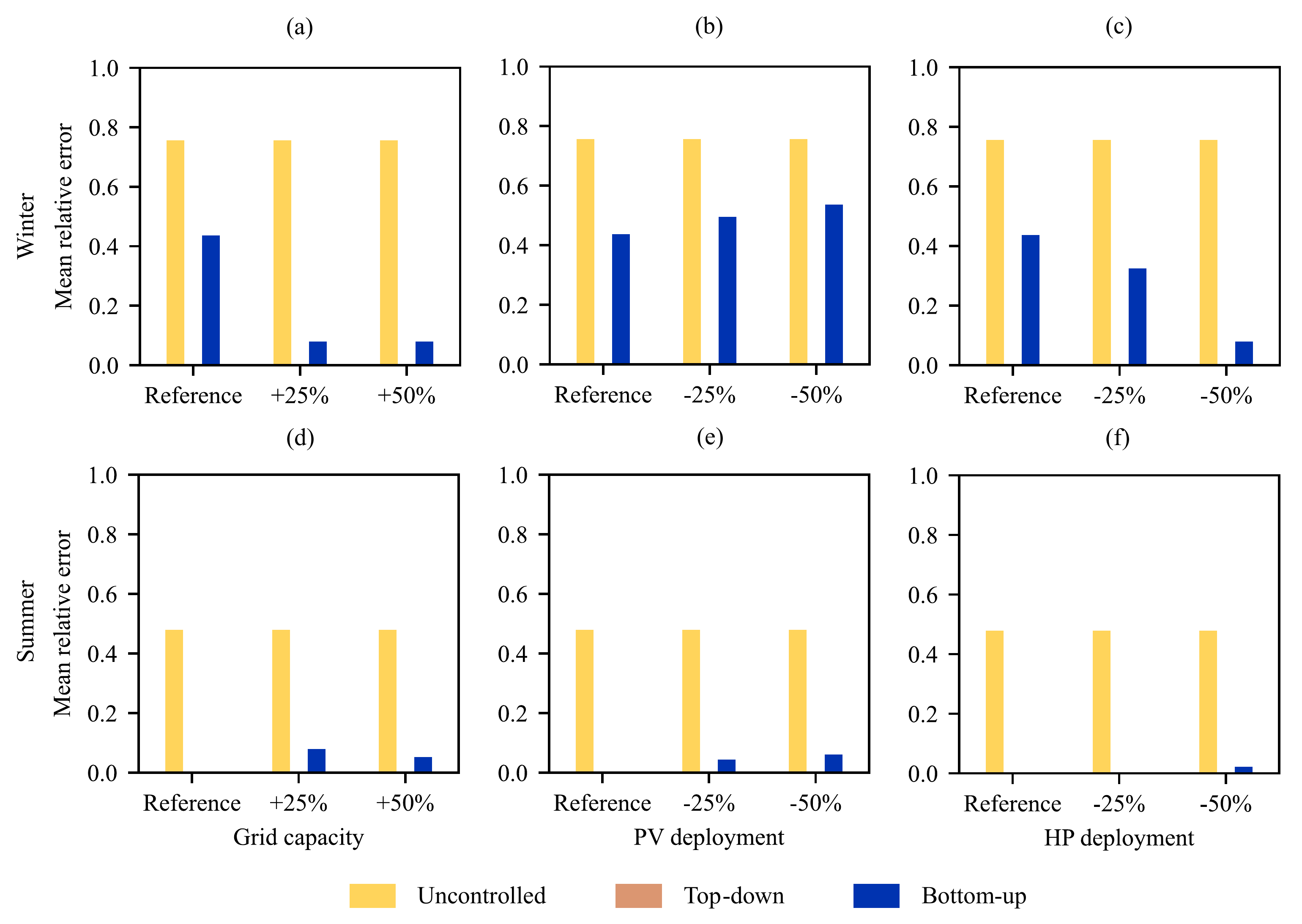}
\caption{Sensitivity analysis results for mean relative deviations of the EV charging approaches from the CP profile, under different experimental settings. The three EV charging approaches (uncontrolled, top-down, and bottom-up) are compared for both the representative winter day (panels (a)--(c)) and the representative summer day (panels (d)--(f)). Panels (a) and (d) show the impact of different levels of grid reinforcement, panels (b) and (e) the impact of PV deployment, and panels (c) and (f) the impact of HP deployment.}
\label{fig:sensitivity_fitting}
\end{figure*}

Figure~\ref{fig:sensitivity_fitting} gives the deviations of the uncontrolled, top-down, and bottom-up profiles from the CP profile.
The results are shown for the \textit{Reference} scenario and for the sensitivity analyses on grid capacity, PV deployment, and HP adoption.
The figure shows that the top-down approach consistently reproduces the CP profile with high accuracy across all sensitivity scenarios, and that overall, the results remain robust across all sensitivity scenarios.

Results for the winter day indicate that increased grid capacity and reduced HP adoption not only reduce grid violations but also lower the mean relative errors.
Conversely, a reduction in PV deployment causes the control objectives of the two approaches to diverge, resulting in larger deviations from the aggregated profile and increased grid violations.
In summer, grid capacity increase and PV reduction provide additional operational flexibility to the distribution system, enabling the bottom-up charging approach to lower the value of its objective function (i.e., costs).
This results in an increased deviation from the CP profile, while maintaining grid violations unchanged with respect to the top-down approach.

\subsection{Interpretation of unresolved grid violations}
Unresolved grid violations occur on both winter and summer days for all EV charging approaches.
However, the summer day exhibits the highest remaining violations, even for the decentralized bottom-up approach.
This indicates that summer operating conditions pose distinct challenges and therefore deserve closer examination.

\begin{figure*}
\centering
\includegraphics[width=2\columnwidth]{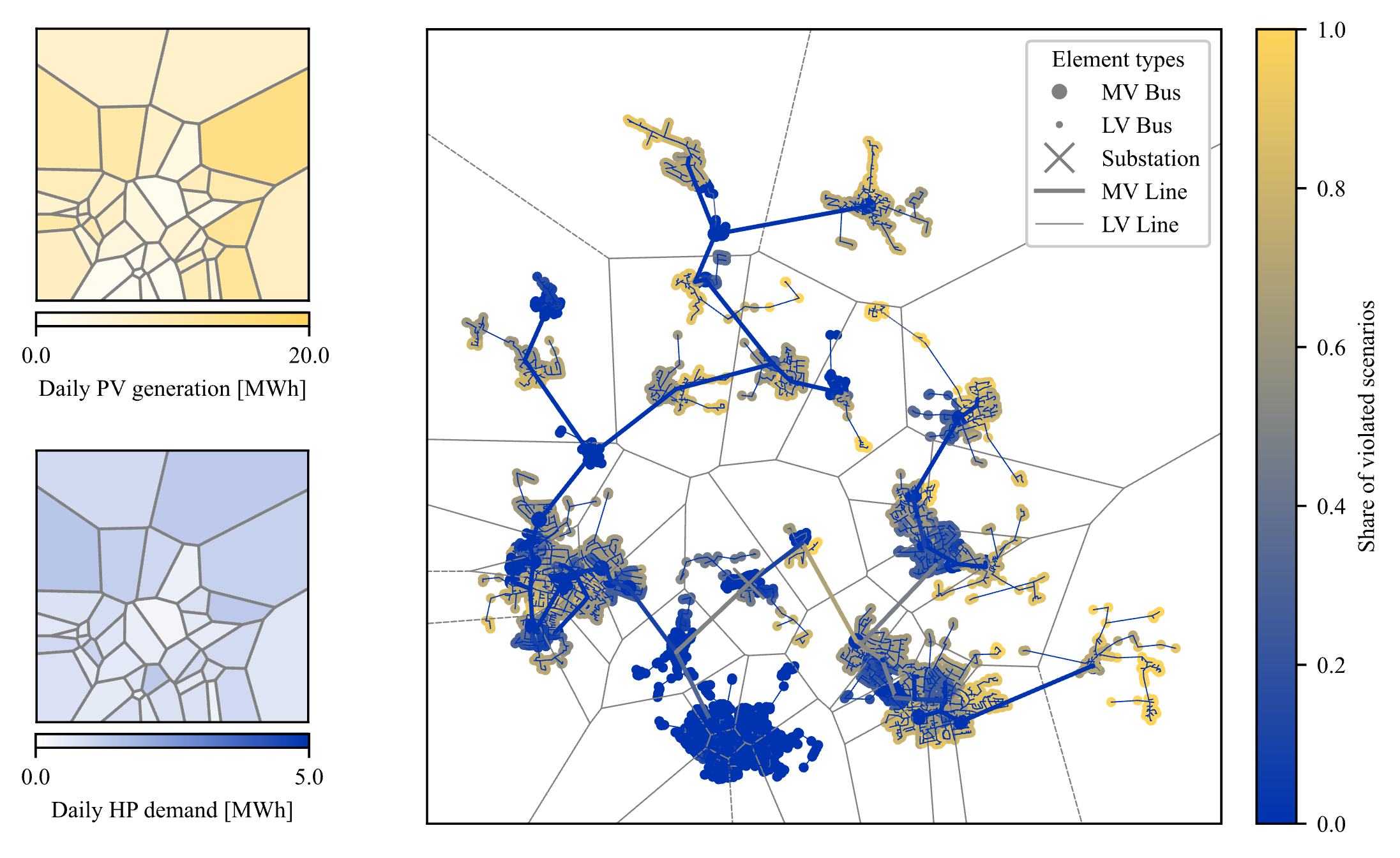}
\caption{Criticality analysis of the integrated medium-voltage and low-voltage grid in terms of operational violations for the representative summer day. The right panel shows the grid topology with buses and lines colored according to the share of scenarios in which they experience a violation (bus voltage or branch flow), with higher shares indicating more critical components. The left panels show the daily PV generation (top) and daily HP demand (bottom) for each LV grid, with the corresponding Voronoi partitions colored.}
\label{fig:violations_summer}
\end{figure*}

Figure~\ref{fig:violations_summer} identifies the critical grid elements (buses and branches) for the bottom-up approach on the summer days.
The criticality of each grid element is quantified as the share of scenarios in which the element experiences a violation during the simulated days, such as a voltage magnitude violation at a bus or a power flow limit violation on a branch.
Here, all simulated summer days are considered, as well as all sensitivity scenarios.
This metric measures the frequency of violations for each component, highlighting the systematically weak elements in the grid.
The figure shows that most violations occur at the LV level.
Additionally, violations are mostly voltage-related and, therefore, occur primarily on peripheral nodes.
Additionally, the left panels of Figure~\ref{fig:violations_summer} show the daily PV generation and HP demand across all LV grids. 
In summer, violations primarily occur in grids with high PV generation, reflecting the stress caused by large PV injections.
The low HP demand, instead, is not influential.

\section{Discussion} 
\label{sec:discussion_conclusion}
This study demonstrates that a CP centrally optimized EV charging demand can deliver substantial benefits not only at the system level but also at the local level. 
This holds even though the CP does not explicitly account for distribution grid constraints or individual driving requirements.
The top-down profile, resulting from the disaggregation of the CP profile to the distribution grid, outperforms uncontrolled charging by reducing grid violations and improving operational feasibility.
This outcome reflects the emerging paradigm of future power systems, where increasingly distributed generation sources are coupled with demand that is both heterogeneous and flexible, leading to greater alignment between central and local grid operational objectives. 
In this study, we demonstrate this paradigm in Switzerland, where plans to electrify transport and heating align with the deployment of rooftop PV systems near these new loads. 

A bottom-up decentralized approach for controlled EV charging can further reduce grid violations compared to a centralized top-down approach.
It does so by identifying an alternative optimal charging schedule that balances the distribution system operator's need to minimize costs and grid violations.
The resulting charging profile remains well aligned with the CP profile, exhibiting similar temporal patterns by concentrating charging demand during daytime hours and reducing the evening peak.
These findings underscore the crucial role of controlled EV charging in mitigating grid stress and deferring costly infrastructure investments, and are consistent with findings from the literature~\cite{Hemmati2020}.

In a system that relies heavily on PV generation, grid operation characteristics differ substantially between winter and summer~\cite{Abideen2019}.
During winter, PV generation and total demand (including conventional, EV, and HP demand) remain relatively well balanced in the considered system.
Consequently, in winter, EV charging flexibility can be effectively leveraged through a bottom-up approach to further reduce grid violations while maintaining a profile that closely matches the time patterns of the top-down profile.
In contrast, in summer, the grid becomes overloaded due to excessive PV generation.
Therefore, during summer, the available EV flexibility is exploited by the bottom-up approach to relieve PV-induced grid constraints, effectively supporting PV integration, and leading the top-down and bottom-up profiles to converge.
These findings underscore that EV flexibility will play a key role in ensuring secure and stable grid operation in the future. Nevertheless, it must be complemented by other flexible resources such as PV curtailment strategies or HP demand flexibility.

The seasonal differences between winter and summer are the underlying cause of the varying outcomes observed in the sensitivity analysis for the winter and summer days.
While increased grid capacity consistently reduces grid violations, the effects of reduced PV deployment and HP adoption differ between winter and summer days.
The winter days are characterized by high demand and low generation, and thus benefit from higher PV injections and lower HP demand. 
In contrast, the summer days are characterized by PV over-generation; consequently, reducing PV injections alleviates grid violations, whereas changes in HP adoption have a negligible impact on system operation. 
Nonetheless, both controlled charging approaches reduce grid violations compared to uncontrolled charging. 
Additionally, the bottom-up approach can further alleviate grid violations by identifying alternative, grid-friendly EV charging schedules.

The results show that the top-down profile does not always match the CP profile.
This is caused by different energy constraints in the CP and top-down models. 
This deviation arises from differences in the temporal energy constraints imposed by the two models, with the CP model enforcing weekly energy balancing and the top-down model enforcing daily energy balancing.

This work comes with a few limitations.
For instance, uncertainty is not considered in the EV driving and charging needs, as in~\cite{Sun2020}.
Both the top-down and bottom-up approaches are based on deterministic models with perfect foresight, and are applied to the analysis of single winter and summer days.
Therefore, to ensure robustness in our findings, we investigate alternative underlying grid conditions, using the same deterministic framework.
Furthermore, we make assumptions about flexibility participation rates, but do not consider the flexibility remuneration. 
Therefore, the underlying assumption is that flexibility providers are always available and willing to participate.
Additionally, since we do not model the single household perspective, in which each user can react to a price signal, we do not capture possible issues of overcoordination identified in the literature~\cite{Schwarz2020}.
Finally, this work focuses exclusively on EV charging flexibility.
While this type of flexibility is shown to be essential for ensuring secure and stable grid operation~\cite{Mastoi2023}, the results also indicate that EV flexibility alone is insufficient to address all future grid challenges.
It should therefore be complemented by other forms of distributed energy resource flexibility~\cite{Bellizio2025}.
Future work could extend the proposed methodology to assess the combined impact of multiple DERs, enabling a more comprehensive understanding of flexibility at the distribution grid level.

\section{Conclusion}
In this work, we present a systematic comparison of three EV charging approaches, namely uncontrolled, top-down, and bottom-up, to assess the impact of centrally optimized EV charging demand on distribution grid operations.
Our findings demonstrate that the centralized top-down approach yields an EV charging dispatch that is typically compatible with individual driving constraints and significantly reduces grid violations compared to uncontrolled charging.
Moreover, the decentralized bottom-up approach can further alleviate grid violations, depending on the underlying grid conditions.
When this is the case, the bottom-up solution features small deviations from the top-down profile, with reduced peak EV demand and more evenly distributed charging demand throughout the day.
The close alignment with the top-down profile underscores the well-suited coordination mechanism based on electricity prices.
Finally, a comprehensive sensitivity analysis on PV deployment, HP adoption, and grid capacity confirms the robustness of these findings across different grid configurations.
To conclude, our answer to this work's title is no, central planning does not fail locally. 
In fact, a centrally optimized EV charging dispatch delivers substantial benefits not only at the system level but also locally.

\section*{CRediT authorship contribution statement}
\textbf{Ambra Van Liedekerke:} Writing – original draft, Investigation, Formal analysis, Conceptualization.
\textbf{Lorenzo Zapparoli:} Writing – review \& editing, Software, Investigation, Formal analysis, Conceptualization.
\textbf{María Parajeles Herrera:} Writing – original draft, Investigation, Formal analysis, Conceptualization.
\textbf{Blazhe Gjorgiev:} Writing – review \& editing, Investigation, Supervision Conceptualization.
\textbf{Gabriela Hug:} Writing – review \& editing, Supervision, Funding acquisition.
\textbf{Giovanni Sansavini:} Writing – review \& editing, Supervision, Conceptualization, Funding acquisition.

\section*{Declaration of Competing Interest}
The authors declare no competing interests.

\section*{Declaration of generative AI and AI-assisted technologies in the manuscript preparation process}
During the preparation of this work, the authors used ChatGPT and Claude in order to improve the text. The authors reviewed and edited the output as needed and take full responsibility for the content of the published article.

\section*{Acknowledgements}
\label{Acknowledgements}
The authors thank Huiwen Luo for the initial work carried out as a Master's thesis.
The research published in this report was carried out with the support of the Swiss Federal Office of Energy SFOE as part of the SWEET consortium EDGE. The authors bear sole responsibility for the conclusions and results.

\bibliographystyle{elsarticle-num-names}
\bibliography{refs}
\end{document}